\documentclass[%
 reprint,
superscriptaddress,
preprintnumbers,
 amsmath,amssymb,
 aps,
]{revtex4-1}

\usepackage{graphicx}
\usepackage{dcolumn}
\usepackage{bm}
\usepackage[colorlinks=true,allcolors=blue]{hyperref}%
\usepackage[flushleft]{threeparttable}
\usepackage{booktabs}
\usepackage{relsize}

\begin{document}


\title{The Anharmonic Origin of the  Giant Thermal Expansion of NaBr}

\author{Y. Shen}
\email{yshen@caltech.edu}
\affiliation{Department of Applied Physics and Materials Science, California Institute of Technology, Pasadena, California 91125, USA}
\author{C. N. Saunders}
\affiliation{Department of Applied Physics and Materials Science, California Institute of Technology, Pasadena, California 91125, USA}
\author{C. M. Bernal}
\affiliation{Department of Applied Physics and Materials Science, California Institute of Technology, Pasadena, California 91125, USA}
\author{D. L. Abernathy}
\affiliation{Neutron Scattering Division, Oak Ridge National Laboratory, Oak Ridge, Tennessee 37831, USA}
\author{M. E. Manley}
\affiliation{Materials Science and Technology Division, Oak Ridge National Laboratory, Oak Ridge, Tennessee 37831, USA}
\author{B. Fultz}
\email{btf@caltech.edu}
\affiliation{Department of Applied Physics and Materials Science, California Institute of Technology, Pasadena, California 91125, USA}

\date{\today}

\begin{abstract}
All phonons in a single crystal of NaBr were measured by
inelastic neutron scattering at temperatures of 10, 300 and 700\;K.
Even at 300\,K the phonons, especially the
longitudinal-optical (LO) phonons, showed
large shifts in frequencies, and 
showed large  broadenings in energy
owing to anharmonicity. 
\emph{Ab initio} computations were first performed
with the quasiharmonic
approximation (QHA), in which
the phonon frequencies depend only on $V$, and on $T$ only 
insofar as it alters $V$ by thermal expansion. 
This QHA was an unqualified failure for predicting the 
temperature dependence of phonon frequencies, even 300\,K,
and the thermal expansion was in error by a factor of four.  
\emph{Ab initio} computations that included both anharmonicity 
and quasiharmonicity successfully predicted both the temperature dependence
of phonons and the large thermal expansion of NaBr. 
The  frequencies of LO phonon modes decrease 
significantly with temperature owing to the real part of the phonon 
self-energy from explicit anhamonicity,
originating from the cubic anharmonicity of nearest-neighbor Na-Br bonds. 
Anharmonicity is not a correction to the QHA predictions of
thermal expansion and thermal phonon shifts, but dominates 
the behavior.
\end{abstract}

\maketitle

Thermal expansion, a fundamental thermophysical property, originates primarily from a competition between the elastic energy of expansion and the phonon entropy, which usually increases beyond harmonic behavior as a solid expands. Thermal expansion can be calculated readily in the quasiharmonic approximation (QHA), which assumes that phonon frequencies depend only on volume \cite{Liu2014, Allen2019, Mittal2018, Nath2016, Grimvall1999, Quong1997}. The QHA theory of thermal expansion is textbook content and is logically self-consistent. It ignores explicit anharmonicity, where phonon frequencies also change with temperature at a fixed volume \cite{Zak2009, Li2011}.
Some calculations include anharmonicity as a small correction to the QHA, but the
relative importance of anharmonicity is  not yet settled \cite{Erba2015, Allen2019}.

We recently found that the QHA gave the wrong sign for the temperature dependence of most phonons in silicon \cite{Kim1992}. This shows that the QHA is physically incomplete, even though it did predict correctly the thermal expansion. 
Here we report a more compelling inelastic neutron scattering (INS)
experiment to test predictions of phonons and thermal expansion in a 
different material, sodium bromide (NaBr). 
Like other alkali halides with the rocksalt structure \cite{Mitskevich1962, Cowley1965, Cowley1968}, NaBr has  received special attention owing to its cubic structure and highly ionic bonding. 

\begin{figure*}
    \centering
    \includegraphics[width=\linewidth]{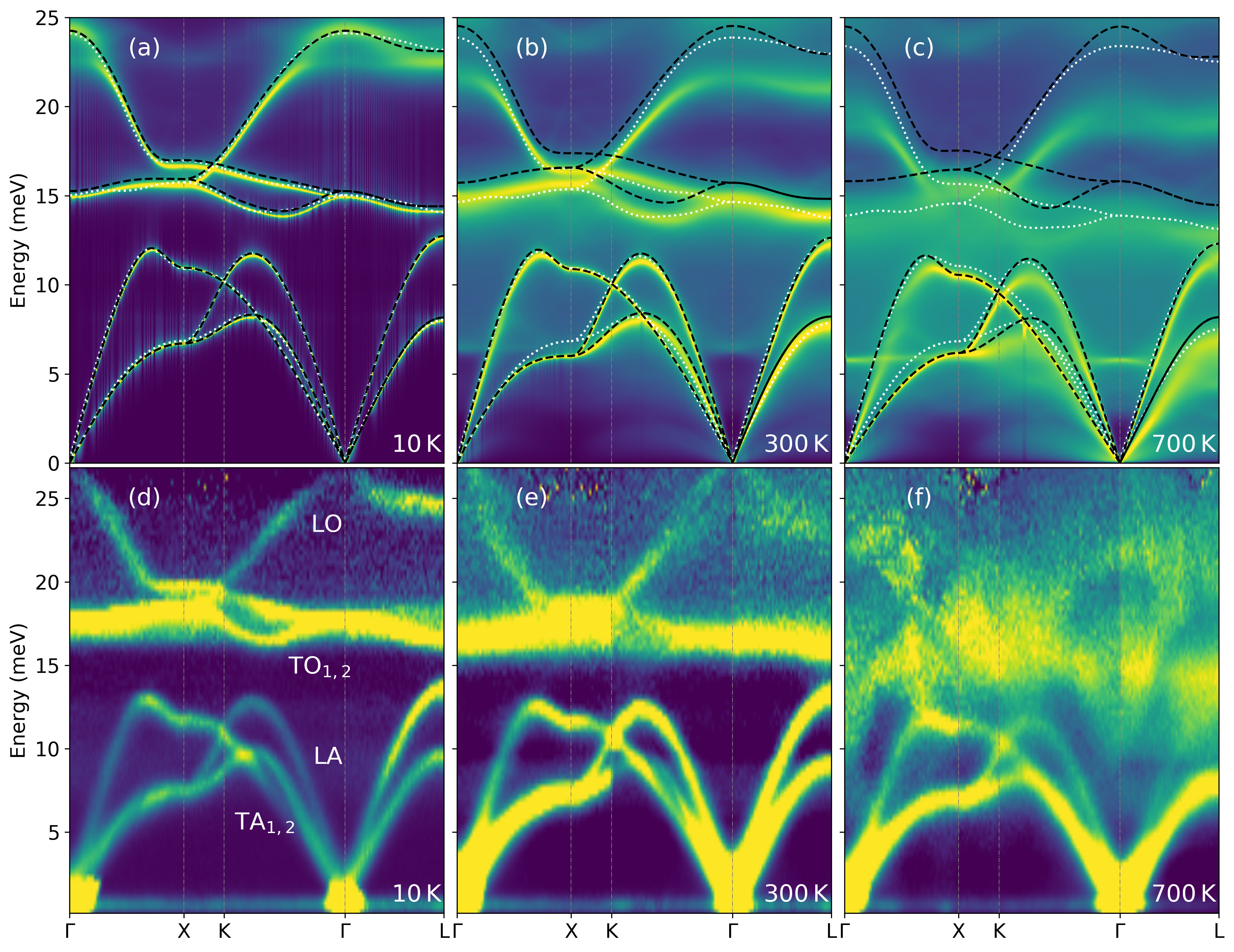}
    \caption{\label{fig: lineshape} Comparison between computational (QHA and fully anharmonic) and experimental (INS) results on phonon dispersions of NaBr. (a-c) Phonons in NaBr calculated with the QHA (white dotted line), with only the second-order force constants from sTDEP (black dashed line), and from the full phonon spectral function (logarithmic intensity map) from sTDEP. Temperatures are labeled in the panels. (d-f) Corresponding 2D slices through the four-dimensional scattering function  $S(\mathbf{Q}, \varepsilon)$, where $\varepsilon = \hbar \omega$, along  high symmetry lines in the first Brillouin zone.
}
\end{figure*}

The INS data from a  single crystal of NaBr
were acquired with the time-of-flight spectrometer, ARCS \cite{Abernathy2012}, at the Spallation Neutron Source at 
the Oak Ridge National Laboratory, using neutrons with an incident energy of 30\,meV. 
Data were collected from 201 rotations of
the crystal in increments of 0.5$^{\circ}$ about the vertical 
[001] axis.
Data reduction gave the 4D scattering function $S(\mathbf{Q}, \varepsilon)$ \cite{Ewing2016, Li2014}, where $\mathbf{Q}$ is the 3D wave-vector and $\varepsilon$ is the phonon energy (from the neutron energy loss). 
Nonlinearities of the ARCS instrument were corrected with a small linear rescaling 
of the $q$-grid, calibrated by the positions of  45 \emph{in situ} Bragg diffractions.
After subtracting the background from measurements on 
an empty can at the same temperature,
and removing multiphonon scattering 
with the incoherent approximation, 
the higher
Brillouin zones  were folded back into an irreducible 
wedge in the first Brillouin zone to obtain the spectral intensities shown in Fig. \ref{fig: lineshape}.
The {\it Supplemental Material} \cite{Sup} describes the experiment and
data analysis in more detail. 

\begin{figure}[!htb]
    \centering
    \includegraphics[width=\columnwidth]{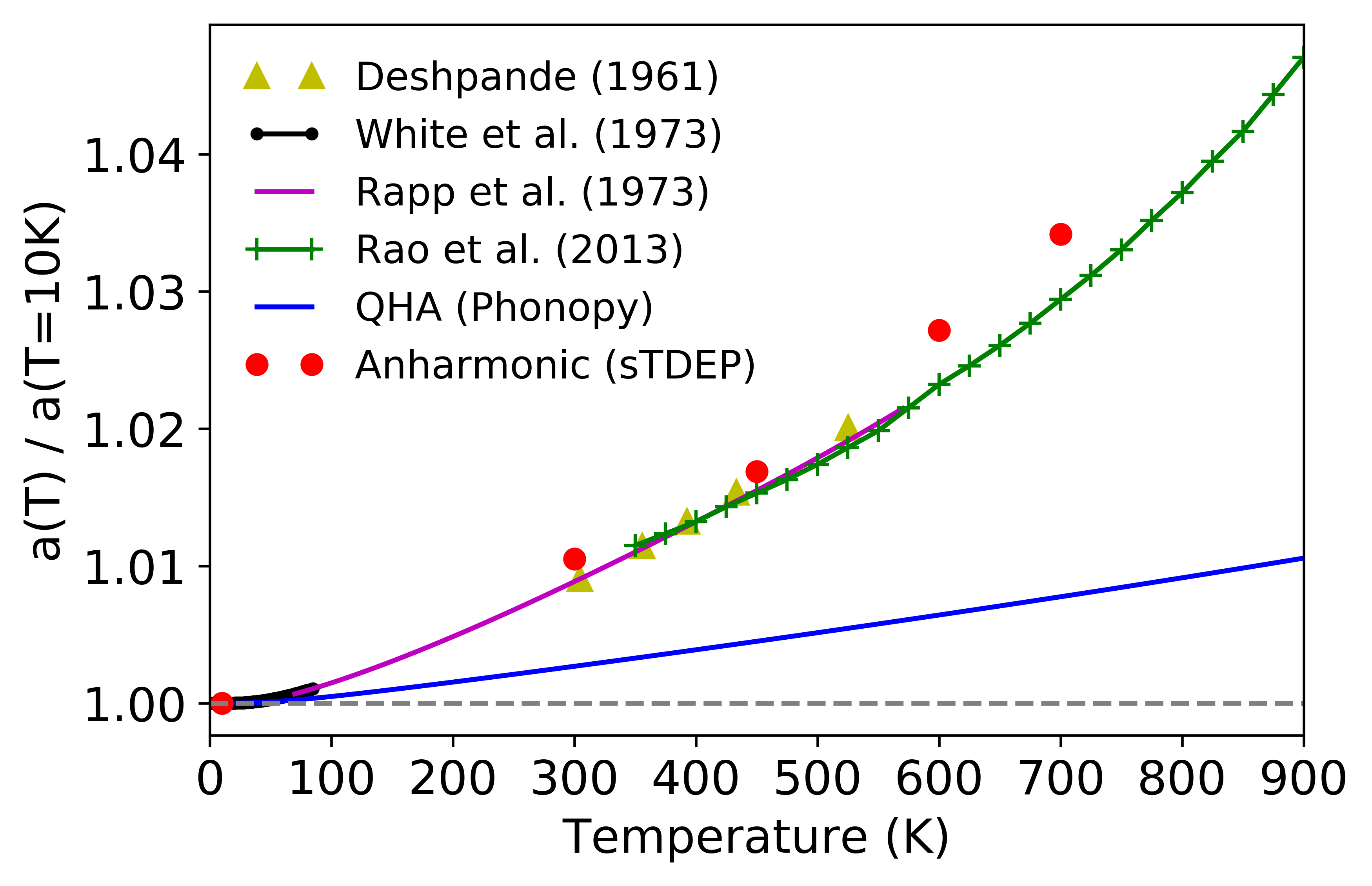}
    \caption{\label{fig:thermal} Thermal expansion of NaBr. The \emph{ab initio} QHA (blue solid line) and anharmonic calculations (red solid circles) are compared with  experimental results \cite{Desh1961, White1973, Rapp1973, Rao2013}. There is a large discrepancy between the measurements and the QHA predictions, while results from the sTDEP method are in close agreement with the experiments.}
\end{figure}

The QHA uses an explicit dependence of phonon frequencies on volume into the Helmholtz free energy 
\begin{eqnarray}
    &{}&F^{\rm QHA}(T, V) \nonumber \\
    &=&U_0(V) + \sum_{\mathbf{q}, j}\left[ \frac{\hbar \omega_{\mathbf{q}, j}}{2}+k_{\rm B}T \ln\left (1- e^{-\frac{\hbar \omega_{\mathbf{q}, j}}{k_{\rm B}T}}\right ) \right] \; ,
\end{eqnarray}
where $U_0(V)$ is the ground-state internal energy without any vibrational contribution and the term $k_{\rm B} \ln[...]$ includes the entropy that depends on volume through the individual phonon frequencies $\omega_{\mathbf{q}, j} = \omega_{\mathbf{q}, j}(V)$ (for the $j$-th phonon branch at wavevector $\mathbf{q}$). 
The finite-displacement method, as implemented in PHONOPY \cite{Togo2015}, was used to obtain phonon frequencies for different volumes by  density functional theory (DFT) calculations with the VASP package \cite{Kresse1993, Kresse1994, Kresse1996, Kresse1996_2}. The equilibrium volume at a given temperature $T$ was obtained by minimizing $F^{\rm QHA}(T, V)$ with respect to volume $V$, keeping $T$ as a fixed parameter. Figure \ref{fig:thermal} shows how the QHA fails to predict both the magnitude and shape of the thermal expansion curve of NaBr, even at room temperature. 

Anharmonic behavior was calculated by the stochastically-initialized temperature dependent effective potential method (sTDEP) \cite{Hellman2011, Hellman2013, Hellman2013_2, Popescu2012}. 
In sTDEP, the Born-Oppenheimer molecular dynamics potential energy surface of NaBr was evaluated by a Monte Carlo sampling of the phase space of atom positions. 
The forces on atoms were fitted to a model Hamiltonian
\begin{eqnarray}
\label{eq: hamiltonian}
    \hat{H}= &U_0&+\sum_i \frac{\textbf{p}_i^2}{2m_i}+
    \frac{1}{2!}\sum_{ij} \sum_{\alpha\beta}\Phi_{ij}^{\alpha\beta}u_i^\alpha u_j^\beta \nonumber\\ 
    &+&\frac{1}{3!}\sum_{ijk} \sum_{\alpha\beta\gamma}\Phi_{ijk}^{\alpha\beta\gamma}u_i^\alpha u_j^\beta u_k^\gamma \; ,
\end{eqnarray}
by DFT calculations on various configurations of displaced atoms by stochastic sampling of a canonical ensemble, with Cartesian displacements ($u_i^\alpha$) normally distributed around the mean thermal displacement. 
The $U_0$ is a fit parameter for the baseline of the potential energy surface. 
The quadratic  constants $\Phi_{ij}$  capture not only harmonic properties, but their temperature dependence accounts for quartic and higher 
nonharmonic parts of the potential. 
These temperature-dependent $\{ \Phi_{ij} \}$ were used to calculate phonon frequencies. 
The cubic force constants $\Phi_{ijk}$ capture 
the broadening and additional shifts of phonon modes, discussed below.

For a given temperature, the Helmholtz free energy $F(T, V)$ was calculated
for different volumes $V$ as \cite{Errea2014}
\begin{eqnarray}
    F(T, V) = U_0(T, V)+F_{\rm vib}(T, V)\; ,
\end{eqnarray}
where $U_0(T, V)$ is the baseline from Eq.~\ref{eq: hamiltonian}. 
The equilibrium volumes were  obtained by minimization of the Helmholtz free energy
at $T$, giving the results shown in  
Fig.~\ref{fig:thermal}.
These equilibrium volumes are in good agreement with 
experimental
measurements, although there are deviations at higher temperatures. 
Details of the calculations of equilibrium volumes and
phonon dispersions are given in the \textit{Supplemental Material} \cite{Sup}.

Some calculated phonon spectral weights are compared to experimental 
measurements in Fig. \ref{fig: lineshape} 
along directions of high symmetry. 
At 10\,K, all calculations agree with each other and with the
experimental measurements. 
At higher temperatures, 
the acoustic dispersions below 14\,meV show some softening, especially at 700\,K,
but are not broadened  so much as  the optical modes.
The optical modes show large broadening at 300\,K, and
major changes in shape at 700\,K. 
The temperature dependence of the optical dispersions is largely captured 
by the spectral weight calculated by sTDEP, but only a minor part of the
softening is predicted by the QHA calculations (and none of the broadening
owing to its assumption of non-interacting modes).
The quadratic term from sTDEP  
(with $\Phi_{ij}^{\alpha\beta}$ in Eq. \ref{eq: hamiltonian}) was used to calculate the dispersions
shown as black dashed lines in Fig.~\ref{fig: lineshape}a-c.
By itself, this term does not reproduce the thermal phonon softening. 
The largest contribution to the temperature shift of the spectral weight is
from the real part of the cubic term, obtained as a Kramers--Kronig transformation of
the imaginary part of the self-energy as explained in the \textit{Supplemental Material} \cite{Sup} with Eq. 51.  
The imaginary part of the phonon self-energy
from this cubic term is responsible for the surprisingly-large energy broadening
of the longitudinal optical (LO) phonons at 300\,K and especially at 700\,K.

\begin{figure*}[!htb]
    \centering
    \includegraphics[width=\linewidth]{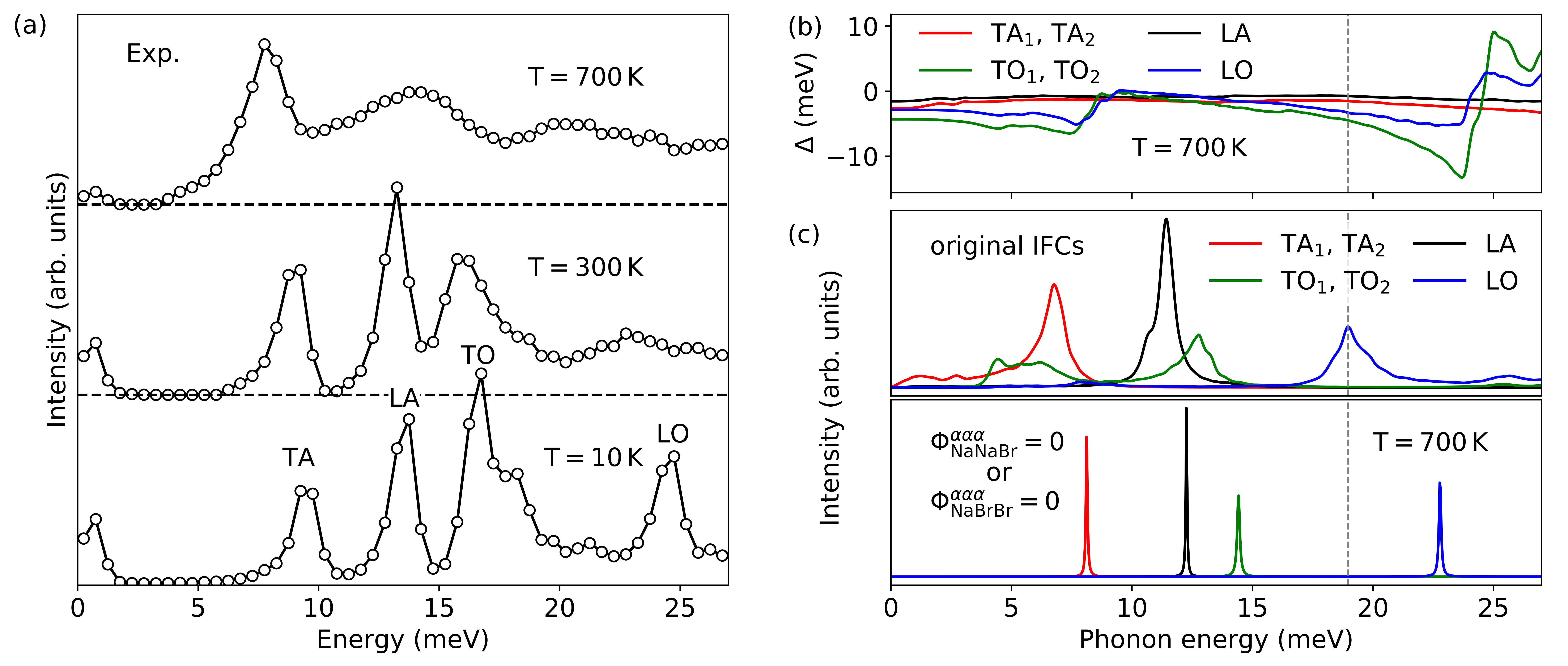}
    \caption{\label{fig: Lpoint} Measured and calculated phonon lineshapes at the $L$-point and the real part of the phonon self-energy. (a) The 1D cut of $S(\mathbf{Q}, \varepsilon)$ at a constant $\mathbf{Q} = [0.5, 0.5, 0.5]\,\mathrm{r.l.u.}$ (reciprocal lattice units), showing the temperature dependence of phonon lineshapes. (The small peak near zero is the residue from elastic scattering after correcting for the phonon creation thermal factor.) (b) Real component of the phonon self-energy $\Delta$ from the third-order force constants. (c) Phonon intensities after nulling the third-order force constants, $\Phi_{\mathrm{NaNaBr}}^{\alpha \alpha \alpha}$ or $\Phi_{\mathrm{NaBrBr}}^{\alpha \alpha \alpha}$, associated with the nearest-neighbor degenerate triplets, where $\alpha = (x, y, z)$ represents the direction along the Na-Br bond.}
\end{figure*}

The experimental inelastic neutron scattering (INS) measurements (see Fig.~\ref{fig: lineshape}) 
were fitted to 
give the energy shifts of LO phonons presented in \hyperref[table: LOshift]{Table I}.
The QHA  accounts for only a  small part of the experimental  shifts, 
but the anharmonic calculations are much more successful. 
The spectral intensities at the $L$-point 
are shown in Fig.~\ref{fig: Lpoint}a. 
All phonons at the $L$-point soften and broaden significantly with temperature. 
Spectra from the longitudinal acoustic (LA) and transverse optic (TO) phonon modes merge into one broad peak at 700\,K. 
The longitudinal optical (LO) peak broadens significantly, 
but its large thermal softening  is still evident.
Figure~\ref{fig: Lpoint}b shows that the real part of the self-energy of the LO phonon 
at the $L$-point is approximately --3.5\,meV at 700\,K,
so phonon-phonon anharmonicity dominates 
the thermal shift of this mode
(the LO mode has a phonon energy of 19\,meV at the $L$-point from sTDEP). 
The \textit{Supplemental Material} \cite{Sup} shows some of the spectral weights in more 
detail. 
There are  differences between  
experiment and the sTDEP calculations at 700\,K, especially halfway between
$\Gamma$ and $L$ between 16 and 23\,meV.
Some anharmonic effects in NaBr are  too large to be predicted accurately
by the sTDEP method.

\addtolength{\tabcolsep}{3pt}
\begin{table*}
  \begin{threeparttable}
    \caption{\textbf{Phonon energy shifts of the LO mode with temperature.}}
     \begin{tabular}{cccccccccc}
        \toprule
        & \multicolumn{9}{c}{Energy shift: $(\varepsilon - \varepsilon_{10\,\rm{K}}) / \varepsilon_{10\,\rm{K}}$} \\
        $T$ (K) & \multicolumn{3}{c}{At $L$-point} & \multicolumn{3}{c}{Along $\Gamma \text{-}L$} & \multicolumn{3}{c}{Along $\Gamma \text{-}X$} \\
        \cmidrule(lr){2-4} \cmidrule(lr){5-7} \cmidrule(lr){8-10}
            & QH & Anh. & Exp. & QH & Anh. & Exp. & QH & Anh. & Exp. \\
        \midrule
        300 & -0.003 & -0.065 & -0.080 (0.020) & -0.037 &  -0.087 &  -0.062 (0.020) & -0.031 & -0.051 &  -0.052 (0.020) \\
        700 & -0.025 & -0.164 & -0.174 (0.055) & -0.045 & -0.181 &  -0.169 (0.055) & -0.034 & -0.144 &  -0.132 (0.055) \\
        \bottomrule
     \end{tabular}
    \begin{tablenotes}
      \small
      \item QH\,=\,quasiharmonic, Anh.\,= anharmonic, Exp.\,=\,experimental.
      \item Errors are from the instrument energy resolution and/or the peak fitting process. 
      \item Average values were used for evaluation along the path.
    \end{tablenotes}
  \end{threeparttable}
\label{table: LOshift}
\end{table*}

To understand the origin of the anharmonicity at 700\,K,
the cubic irreducible force constants (IFCs) 
for the three-body interactions
within the first ten coordination shells were individually set
to zero while recalculating phonon lineshapes at different $\mathbf{Q}$. 
Figure~\ref{fig: Lpoint}c shows how two related IFCs dominate the lineshapes. 
They correspond to the nearest-neighbor cubic interactions of degenerate triplets 
(NaNaBr and/or NaBrBr) in the [100] direction (i.e. along the Na-Br bond direction).
(By translational invariance, $\Phi_{\rm NaNaBr}^{\alpha \alpha \alpha}  = - \Phi_{\rm NaBrBr}^{\alpha \alpha \alpha}$.) 
When these force constants are switched off, 
the phonon lineshapes revert to narrow Lorentzian functions
typical of weakly anharmonic solids, and these Lorentzian peaks
are at  energies similar to those from  the QHA calculations. 
The dominance of $\Phi_{\rm NaNaBr}^{\alpha \alpha \alpha}  = - \Phi_{\rm NaBrBr}^{\alpha \alpha \alpha}$
on the phonon anharmonicity 
was  found for phonons at all other points in  reciprocal space, as shown in the \textit{Supplemental Material} \cite{Sup}. 

The physics of thermal expansion
requires volume and temperature derivatives of $F(V,T)$, 
specifically $\partial ^2 F /\left( \partial V \, \partial T\right) = - {\beta \, B_T}$.
The \textit{Supplemental Material} \cite{Sup} obtains an expression for the ratio between $\beta^{\rm QH}$, the thermal expansion in the QHA, and the real $\beta$. For $\hbar \omega_{\rm max} < k_{\rm B}T$, 
\begin{align}
    \beta^{\rm QH}\Big/\beta = 1 - \frac{6k_{\rm B}}{\beta \, B_T \, v} \, \mathlarger{\overline{\gamma}}_{\mathsmaller{V, T}} \; ,
\end{align}
where $B_T$ is the isothermal bulk modulus, $v$ is the volume per atom, the mode anharmonicity parameter is
\begin{equation}
    \mathlarger{\gamma}_{\mathsmaller{V, T}}\triangleq -\frac{VT}{\omega}\left ( \frac{\partial^2 \omega}{\partial T\partial V}\right ) \; ,
\end{equation}
and $\mathlarger{\overline{\gamma}}_{\mathsmaller{V, T}}$ is the average anharmonicity parameter. For NaBr, $\beta^{\rm QH} \simeq 0.28\beta$, which is consistent with Fig. \ref{fig:thermal} above. 

By testing different first-principles calculations against
phonons measured by inelastic neutron scattering  at different temperatures, 
we  demonstrated that the widely-accepted quasiharmonic method 
predicts only a small fraction 
of the thermal phonon shifts and the thermal expansion.
Anharmonic effects drastically alter the phonon self energies, especially
the LO phonons. 
The dominant anharmonicity is from cubic interactions associated with the nearest-neighbor degenerate triplets along the Na-Br boding direction. 
The volume dependence of the 
phonon anharmonicity dominates  the thermal expansion of NaBr, and may do so in many other materials.

\bigskip

We thank D. Kim, O. Hellman, F. Yang and J. Lin for helpful discussions. Research with the Spallation Neutron Source (SNS) at the Oak Ridge National Laboratory was sponsored by the Scientific User Facilities Division, Basic Energy Sciences (BES), Department of Energy (DOE). This work used resources from National Energy Research Scientific Computing Center (NERSC), a DOE Office of Science User Facility supported by the Office of Science of the US Department of Energy under Contract DE-AC02-05CH11231. This work was supported by the DOE Office of Science, BES, under Contract DE-FG02-03ER46055.

\providecommand{\noopsort}[1]{}\providecommand{\singleletter}[1]{#1}%
%


\end{document}


\title{Supplementary Material for:\\ The Anharmonic Origin of the Giant Thermal Expansion of NaBr}

\author{Y. Shen}
\email{yshen@caltech.edu}
\affiliation{Department of Applied Physics and Materials Science, California Institute of Technology, Pasadena, California 91125, USA}
\author{C. N. Saunders}
\affiliation{Department of Applied Physics and Materials Science, California Institute of Technology, Pasadena, California 91125, USA}
\author{C. M. Bernal}
\affiliation{Department of Applied Physics and Materials Science, California Institute of Technology, Pasadena, California 91125, USA}
\author{D. L. Abernathy}
\affiliation{Neutron Scattering Division, Oak Ridge National Laboratory, Oak Ridge, Tennessee 37831, USA}
\author{M. E. Manley}
\affiliation{Material Science and Technology Division, Oak Ridge National Laboratory, Oak Ridge, Tennessee 37831, USA}
\author{B. Fultz}
\email{btf@caltech.edu}
\affiliation{Department of Applied Physics and Materials Science, California Institute of Technology, Pasadena, California 91125, USA}

\date{\today}

\maketitle
\section{Comparing the thermal expansion in quasiharmonic and anharmonic theories\label{FirstSection}}

\subsection{Classical Thermodynamics}

The (volumetric) thermal expansion coefficient, $\beta$, is defined as
\begin{equation}
    \beta = \frac{1}{V}\frac{{\rm d}V}{{\rm d}T} \; 
    \label{TherExpDefinition}
\end{equation}
at $P=0$, or at a constant pressure.
This Section \ref{FirstSection} obtains $\beta$ from 
the thermodynamic free energy $F(V,T)$,
which includes the  variables of Eq. \ref{TherExpDefinition}. 
A good starting point is  classical thermodynamics, which
provides relationships 
involving partial derivatives of $F$ with respect to $V$ and $T$.
Thermal expansion requires both types of partial derivatives 
to second order. 
Expansion is a change in volume, of course, but thermal expansion  occurs
with a change of temperature. 

We start with a thermodynamic identity
\begin{equation}
    \beta = -\frac{1}{B_T}\frac{\partial^2 F }{\partial T \partial V} \; ,
    \label{identityTherm}
\end{equation}
where $B_T$ is the isothermal bulk modulus, defined as
\begin{equation}
    B_T = -V\frac{\partial P}{\partial V} 
    = V\frac{\partial^2 F}{\partial V^2} \; .
\end{equation}
Thus, we have 
\begin{align}
    \beta =  -\frac{1}{V} \left(\frac{\partial^2 F }{\partial T \partial V}\right) \Big/ \left(\frac{\partial^2 F }{\partial V^2}\right) \; .
    \label{eq: gen_beta}
\end{align}

One strategy to calculate 
thermal expansion is to solve for
$\displaystyle f(T, V)\triangleq \frac{\partial F}{\partial V}=0$ for $V=V(T)$ in  
quasiharmonic or anharmonic models, and then obtain the thermal expansion coefficient from 
Eq.~\ref{TherExpDefinition}. 
This is done for the quasiharmonic approximation
in Sect. \ref{ab initio calculations} with Fig. \ref{fig: sup_qha}.
Here we  employ an alternative strategy of calculating Eq.~\ref{eq: gen_beta} directly. The two approaches are equivalent because
\begin{align}
    \frac{{\rm d}V}{{\rm d}T} 
    = -\left(\frac{\partial f}{\partial T}\right) \Big/ \left(\frac{\partial f}{\partial V}\right)
    = -\left(\frac{\partial^2 F }{\partial T \partial V}\right) \Big/ \left(\frac{\partial^2 F }{\partial V^2}\right) \; .
\end{align}

\subsection{Phonon Statistical Mechanics}

We address the underlying physics by calculating the phonon free energy,
ignoring possible contributions from electronic or magnetic excitations. 
The key quantities are the energies $\{ \hbar \omega_i \}$, where 
$\omega_i$ is the frequency of a phonon added to the $i$th vibrational mode.
In general, this frequency depends on $V$ and $T$, $\omega_i (V,T)$.
In the  ``quasiharmonic approximation'' (QHA), the $\omega_i$ and the phonon free energy
depend only on $V$. 
In the QHA, the effects of temperature are only from the occupancies of  phonon modes. 
The QHA is convenient for calculating thermal expansion, requiring only a set of  
mode Gr\"{u}neisen parameters. 
It was recently proved that the QHA gives the leading term of the quantity ${\partial F} / {\partial V}$ \cite{Allen2019}, also \cite{Wallace2002}. 
This does not guarantee that the QHA gives the leading term of the thermal expansion coefficient, however,
because a temperature derivative is still needed. (For example, consider the functional $y=y_0+\Delta y $, 
where 
$y_0= 100 + 0.001 \sin x$ and $\Delta y=\sin x$. 
Here the $x$-derivative is dominated by the small term $\Delta y$, rather than the leading term $y_0$.)

Less well known are anharmonic theories of thermal expansion 
developed with many-body theory \cite{Cowley1967, Reissland1973}.
To our knowledge, there has been no direct comparison
of thermal expansion from the QHA and anharmonic theory,
so the goal of this section is an comparison of the relative importance
of each. 

Ignoring the entropy from free electrons or magnetic excitations, for a simple solid with $N$ atoms the Helmholtz free energy originates with the  electronic energy, i.e.,  the internal energy of
the lattice, $U_0$,
plus the free energy from phonons
\begin{align}
    F &=   U_0 + F_{\rm ph} \; , \nonumber \\
    F &= U_0 + \left \langle\sum_{\mathbf{k}, j}\left [\frac{1}{2}\hbar \omega_{\mathbf{k}, j} + k_{\rm B}T\ln{\left (1 - e^{-\frac{\hbar \omega_{\mathbf{k}, j}}{k_{\rm B}T}}\right )}\right ]\right \rangle_{\mathrm{BZ}} \; .
\end{align}
where $\hbar \omega_{\mathbf{k}, j}$ is the phonon energy at the $\mathbf{k}$-point for the $j$th phonon branch.
The sum is taken over all phonon branches and $\mathbf{k}$-points in  reciprocal space, and $\left \langle...\right \rangle_{\mathrm{BZ}}$ is the average over the first Brillouin zone. 
For clarity in what follows, the subscripts of $\omega$ are suppressed. 

Because the free energy depends on $V$ and $T$, we expect $U_0=U_0(V, T)$ and $\omega=\omega(V, T)$.
Derivatives of these
quantities are needed for Eq. \ref{eq: gen_beta}.
The volume derivative is essential
for expansion
\begin{align}
    \frac{\partial F}{\partial V}
    &= \frac{\partial U_0}{\partial V} + \left \langle\sum \frac{\hbar}{2}\frac{\partial \omega}{\partial V}+\frac{k_{\rm B}T}{1-e^{-\frac{\hbar \omega}{k_{\rm B}T}}}\left ( -e^{-\frac{\hbar \omega}{k_{\rm B}T}}\right )\left (-\frac{\hbar}{k_{\rm B}T}\right )\frac{\partial \omega}{\partial V} \right \rangle_{\mathrm{BZ}} \nonumber \\
    &= \frac{\partial U_0}{\partial V} + \frac{\hbar}{2}\left \langle\sum\frac{\partial \omega}{\partial V}\coth{\left( \frac{\hbar \omega}{2k_{\rm B}T}\right )} \right \rangle_{\mathrm{BZ}} \; .
    \label{eq: fv}
\end{align}
From Eq.~\ref{eq: fv}, we can calculate
the additional derivatives needed in 
Eq. \ref{eq: gen_beta}
\begin{align}
    \frac{\partial^2 F }{\partial T \partial V} = \frac{\partial ^2 U_0}{\partial T\partial V} + \frac{\hbar}{2}\left \langle\sum \left[\frac{\partial ^2 \omega}{\partial T\partial V}\coth{\left( \frac{\hbar \omega}{2k_{\rm B}T}\right)} - \frac{\partial \omega}{\partial V} 	\operatorname{csch}^2 \left( \frac{\hbar \omega}{2k_{\rm B}T}\right) \left( \frac{\hbar}{2k_{\rm B}T}\frac{\partial \omega}{\partial T}-\frac{\hbar \omega}{2k_{\rm B}T^2}\right)\right]\right \rangle_{\mathrm{BZ}} \; ,
    \label{eq: FTV}
\end{align}
and
\begin{align}
    \frac{\partial^2 F }{\partial V^2} = \frac{\partial ^2 U_0}{\partial V^2} + \frac{\hbar}{2}\left \langle\sum \left[\frac{\partial ^2 \omega}{\partial V^2}\coth{\left( \frac{\hbar \omega}{2k_{\rm B}T}\right)} - \frac{\partial \omega}{\partial V} 	\operatorname{csch}^2 \left( \frac{\hbar \omega}{2k_{\rm B}T}\right) \left( \frac{\hbar}{2k_{\rm B}T}\frac{\partial \omega}{\partial V}\right)\right]\right \rangle_{\mathrm{BZ}} \; .
    \label{eq: FVV}
\end{align}

\subsection{Internal Energy}

The first terms with $U_0$ in Eqs. \ref{eq: FTV} and
\ref{eq: FVV} are familiar from the
elastic energy of a solid.
They are typically obtained from a          Taylor expansion of the internal energy.
For cubic crystals near the ground-state equilibrium volume $V_0$,
            \begin{align}
                &\qquad U_0(V, T) = U_0(V_0, T)+\frac{B_0(T)V_0}{2}\left (\frac{V-V_0}{V_0}\right )^2+... \; ,
            \end{align}
            gives
            \begin{align}
                \frac{\partial U_0}{\partial V} = B_0(T)\left( \frac{V-V_0}{V_0} \right) \; ,
            \end{align}
            then 
            \begin{equation}
                \frac{\partial^2 U_0}{\partial V^2} = \frac{B_0(T)}{V_0} \; ,
            \end{equation}
            and
            \begin{equation}
                 \frac{\partial^2 U_0}{\partial T\partial V} = \frac{dB_0}{dT}\left( \frac{V-V_0}{V_0} \right) \; ,
            \end{equation}
        where $B_0=-V(\partial P/\partial V)_{P=0}$ is the zeroth-order bulk modulus (i.e. $B_0=B_T$).

\subsection{Phonon Contributions at Medium to High Temperatures\label{AnhSubSection}}

To simplify the second terms in 
Eqs. \ref{eq: FTV} and
\ref{eq: FVV}, notice that $\displaystyle \coth{(x)}=\frac{1}{x}+\frac{x}{3}-\frac{x^3}{45}+\ldots$ and $\displaystyle \operatorname{csch}(x)=\frac{1}{x}-\frac{x}{6}+\frac{7x^3}{360}+\ldots$, so for $\hbar \omega_{\rm max} < k_{\rm B}T$, 
            \begin{equation}
                \coth{\left( \frac{\hbar \omega}{2k_{\rm B}T}\right)} \simeq \operatorname{csch}\left( \frac{\hbar \omega}{2k_{\rm B}T}\right) \simeq \frac{2k_{\rm B}T}{\hbar \omega} \;. \label{ApproxCothCoth}
            \end{equation}
Using Eq. \ref{ApproxCothCoth} greatly
simplifies Eq.~\ref{eq: FTV} and \ref{eq: FVV}, with the restriction to  higher temperatures
where $\hbar \omega_{\rm max} < k_{\rm B}T$,
\begin{align}
    \frac{\partial^2 F }{\partial T \partial V} 
    &= \frac{dB_T}{dT}\left( \frac{V-V_0}{V_0} \right) + k_{\rm B}\left \langle\sum \left(\frac{T}{\omega} \frac{\partial ^2 \omega}{\partial T\partial V} - \frac{T}{\omega^2}\frac{\partial \omega}{\partial T} \frac{\partial \omega}{\partial V}+\frac{1}{\omega} \frac{\partial \omega}{\partial V} \right)\right \rangle_{\mathrm{BZ}}\; ,
    \label{eq: FTV2}
\end{align}
and
\begin{align}
    B_T = V\frac{\partial^2 F }{\partial V^2} 
    &\simeq V\left\{ \frac{B_T}{V_0} + k_{\rm B}T\left \langle\sum \left[\frac{1}{\omega} \frac{\partial ^2 \omega}{\partial V^2} - \frac{1}{\omega^2} \left(\frac{\partial \omega}{\partial V}\right)^2\right]\right \rangle_{\mathrm{BZ}}\right\} \nonumber \\
    B_T  &= \frac{V}{V_0}B_T + k_{\rm B}TV\left \langle\sum \frac{\partial ^2 (\ln{\omega})}{\partial V^2}\right \rangle_{\mathrm{BZ}}\; ,
    \label{eq: FVV2}
\end{align}
which indicates $\displaystyle \left|k_{\rm B}TV\left \langle\sum \frac{\partial ^2 (\ln{\omega})}{\partial V^2}\right \rangle_{\mathrm{BZ}}\right|\ll B_T$, since $V/V_0\simeq 1$.

\subsection{Quasiharmonic Approximation\label{QHSubSection}}

In the quasiharmonic 
approximation, $U_0=U_0(V, T=0)$ and $\omega=\omega(V, T=0)$, so
\begin{align}
    \displaystyle \left(\frac{\partial^2 U_0}{\partial V^2}\right)^{\rm QH} = \frac{B_T(T=0)}{V_0}\; , \quad \displaystyle \left(\frac{\partial^2 U_0}{\partial T\partial V}\right)^{\mathrm{QH}}=0 \; , \quad \displaystyle \left(\frac{\partial \omega}{\partial T}\right)^{\mathrm{QH}}=0 \; .
\end{align}
For $\hbar \omega_{\rm max} < k_{\rm B}T$, Eqs.~\ref{eq: FTV} and \ref{eq: FVV} are
simplified
\begin{align}
    \left(\frac{\partial^2 F }{\partial T \partial V}\right)^{\rm QH} 
    &= k_{\rm B}\left \langle\sum \frac{1}{\omega(V, T=0)} \frac{\partial \omega(V, T=0)}{\partial V}\right \rangle_{\mathrm{BZ}}\; ,
    \label{eq: FTV2QH}
\end{align}
and
\begin{align}
    B_T^{\rm QH} = V\left(\frac{\partial^2 F }{\partial V^2}\right)^{\rm QH} \simeq \frac{V}{V_0}B_T(T=0) + k_{\rm B}TV\left \langle\sum \frac{\partial ^2 \left[\ln{\omega(V, T=0)}\right]}{\partial V^2}\right \rangle_{\mathrm{BZ}}\simeq B_T(T=0)\simeq B_T\; ,
    \label{eq: FVV2QH}
\end{align}
assuming $B_T$ is not strongly dependent on temperature, which is often true in practice.

\subsection{Comparison of Quasiharmonic
and Anharmonic Results for Thermal Expansion}

The results from Sections \ref{AnhSubSection} and \ref{QHSubSection}
allow a direct comparison
of the difference in thermal expansion
predicted by anharmonic and quasiharmonic
theory. 
For moderate to high temperatures
the difference between $\beta$ and $\beta^{\rm QH}$ is 
\begin{align}
    \beta -\beta^{\rm QH}
    &= \left(-\frac{1}{B_T}\frac{\partial^2 F }{\partial T \partial V}\right)-\left(-\frac{1}{B_T^{\rm QH}}\frac{\partial^2 F^{\rm QH} }{\partial T \partial V}\right) \; , \nonumber \\
    &\simeq -\frac{1}{B_T} \frac{\partial^2\left(F-F^{\rm QH}\right)}{\partial T \partial V} \nonumber \; ,
    \end{align}
assuming the bulk modulus does
not vary strongly with temperature.
Using the derivatives of $F$ from
Sections \ref{AnhSubSection} and \ref{QHSubSection}
\begin{align}
    \beta -\beta^{\rm QH}
    &= -\frac{1}{B_T}\frac{dB_T}{dT}\left( \frac{V-V_0}{V_0} \right) - \nonumber \\
    &\quad \frac{k_{\rm B}}{B_T}\left \langle\sum \left(\frac{T}{\omega} \frac{\partial ^2 \omega}{\partial T\partial V} - \frac{T}{\omega^2}\frac{\partial \omega}{\partial T} \frac{\partial \omega}{\partial V}+\frac{1}{\omega(V, T)} \frac{\partial \omega(V, T)}{\partial V} - \frac{1}{\omega(V, T=0)} \frac{\partial \omega(V, T=0)}{\partial V}\right)\right \rangle_{\mathrm{BZ}} \nonumber \\
    &\simeq  -\frac{1}{B_T}\frac{dB_T}{dT}\left( \frac{V-V_0}{V_0} \right) - \frac{k_{\rm B}}{B_T}\left \langle\sum \left[\frac{T}{\omega} \frac{\partial ^2 \omega}{\partial T\partial V} - \frac{T}{\omega^2}\frac{\partial \omega}{\partial T} \frac{\partial \omega}{\partial V}+T\frac{\partial}{\partial T}\left( \frac{1}{\omega}\frac{\partial \omega}{\partial V}\right)\right]\right \rangle_{\mathrm{BZ}} \nonumber \\
    &=  -\frac{1}{B_T}\frac{dB_T}{dT}\left( \frac{V-V_0}{V_0} \right) - \frac{2k_{\rm B}}{B_T}\left \langle\sum \left(\frac{T}{\omega} \frac{\partial ^2 \omega}{\partial T\partial V} - \frac{T}{\omega^2}\frac{\partial \omega}{\partial T} \frac{\partial \omega}{\partial V}\right)\right \rangle_{\mathrm{BZ}} \nonumber \\
    &=  -\frac{1}{B_T}\frac{dB_T}{dT}\left( \frac{V-V_0}{V_0} \right) + \frac{2k_{\rm B}T}{B_T}\left \langle\sum \left(-\frac{1}{\omega} \frac{\partial ^2 \omega}{\partial T\partial V} + \frac{1}{\omega^2}\frac{\partial \omega}{\partial T} \frac{\partial \omega}{\partial V}\right)\right \rangle_{\mathrm{BZ}}\; ,
\end{align}
thus we have
\begin{align}
    \beta^{\rm QH} \Big/ \beta 
    &= 1 + \frac{1}{B_T\beta}\frac{dB_T}{dT}\left( \frac{V-V_0}{V_0} \right) - \frac{2k_{\rm B}T}{B_T\beta}\left \langle\sum \left(-\frac{1}{\omega} \frac{\partial ^2 \omega}{\partial T\partial V} + \frac{1}{\omega^2}\frac{\partial \omega}{\partial T} \frac{\partial \omega}{\partial V}\right)\right \rangle_{\mathrm{BZ}} \nonumber \\
    &\simeq 1 - \frac{2k_{\rm B}T}{B_T\beta}\left \langle\sum \left(-\frac{1}{\omega} \frac{\partial ^2 \omega}{\partial T\partial V} + \frac{1}{\omega^2}\frac{\partial \omega}{\partial T} \frac{\partial \omega}{\partial V}\right)\right \rangle_{\mathrm{BZ}}\; ,
    \label{eq: ratio}
\end{align}
since $\displaystyle \frac{1}{B_T\beta}\frac{dB_T}{dT}\left( \frac{V-V_0}{V_0} \right) < \frac{1}{B_T\beta}\frac{dB_T}{dT} \beta T=\left( \frac{dB_T}{dT}T\right) \Big/ B_T\sim 0.1 \ll 1$ (e.g. in MgO \cite{Anderson1966}).

\subsection{Simplifying Parameters}

Introducing the mode Gr$\ddot{\rm u}$neisen parameter,
\begin{equation}
    \mathlarger{\gamma}_{\mathsmaller{V}}\triangleq -\frac{V}{\omega}\left ( \frac{\partial \omega}{\partial V}\right )\bigg|_T \; , 
\end{equation} 
the thermal Gr$\ddot{\rm u}$neisen parameter (unitless), 
\begin{equation}
    \mathlarger{\gamma}_{\mathsmaller{T}}\triangleq -\frac{T}{\omega}\left ( \frac{\partial \omega}{\partial T}\right )\bigg|_V \; , 
\end{equation} 
and the anharmonicity parameter (unitless),
\begin{equation}
    \mathlarger{\gamma}_{\mathsmaller{V, T}}\triangleq -\frac{VT}{\omega}\left ( \frac{\partial^2 \omega}{\partial T\partial V}\right ) \; ,
    \label{AnhParameterDefined}
\end{equation} 
Eq.~\ref{eq: ratio} can be rewritten as 
\begin{align}
    \beta^{\rm QH}\Big/\beta = 1 - \frac{2k_{\rm B}}{B_T\beta V}\left \langle\sum \left( \mathlarger{\gamma}_{\mathsmaller{V, T}} + \mathlarger{\gamma}_{\mathsmaller{T}} \mathlarger{\gamma}_{\mathsmaller{V}} \right)\right \rangle_{\mathrm{BZ}} \; .
\end{align}

To first order, the infinitesimal change of phonon energy can be divided into changes induced by temperature and volume
\begin{equation}
    \Delta \omega \simeq \frac{\partial \omega}{\partial T} \Delta T + \frac{\partial \omega}{\partial V} \Delta V \triangleq \Delta \omega^{\rm T}+\Delta \omega^{\rm V} \; ,
\end{equation} 
we find (for later use in Eq. \ref{eq: ratio})
\begin{align}
    -\frac{1}{\omega} \frac{\partial ^2 \omega}{\partial T\partial V} + \frac{1}{\omega^2}\frac{\partial \omega}{\partial T} \frac{\partial \omega}{\partial V} 
    &\simeq -\frac{1}{\omega}\frac{\Delta \omega^{\rm T}+\Delta \omega^{\rm V}}{\Delta T \Delta V} + \frac{1}{\omega^2} \frac{\Delta \omega^{\rm T}}{\Delta T} \frac{\Delta \omega^{\rm V}}{\Delta V} \nonumber \\
    &= \frac{1}{\Delta T \Delta V}\left[ \left(-\frac{\Delta \omega^{\rm T}}{\omega}\right) + \left(- \frac{\Delta \omega^{\rm V}}{\omega}\right) + \left(-\frac{\Delta \omega^{\rm T}}{\omega}\right) \left(-\frac{\Delta \omega^{\rm V}}{\omega}\right)\right] \; .
\end{align}
For small positive values of $\displaystyle \left(-\frac{\Delta \omega^{\rm T}}{\omega}\right)$ and $\displaystyle \left(-\frac{\Delta \omega^{\rm V}}{\omega}\right)$,
\begin{equation}
    \left(-\frac{\Delta \omega^{\rm T}}{\omega}\right) + \left(- \frac{\Delta \omega^{\rm V}}{\omega}\right) \gg \left(-\frac{\Delta \omega^{\rm T}}{\omega}\right) \left(-\frac{\Delta \omega^{\rm V}}{\omega}\right) \;, 
\end{equation}
which gives
\begin{align}
    \beta^{\rm QH}\Big/\beta \simeq 1 - \frac{2k_{\rm B}}{B_T\beta V}\left \langle\sum \mathlarger{\gamma}_{\mathsmaller{V, T}} \right \rangle_{\mathrm{BZ}} = 1 - \frac{2k_{\rm B}}{B_T\beta V}(3N)\mathlarger{\overline{\gamma}}_{\mathsmaller{V, T}} = 1 - \frac{6k_{\rm B}}{B_T\beta v}\mathlarger{\overline{\gamma}}_{\mathsmaller{V, T}}\; ,
    \label{eq: ratio_final}
\end{align}
where $v = {V}/{N}$ is the volume per atom, and $\displaystyle \mathlarger{\overline{\gamma}}_{\mathsmaller{V, T}} = \frac{1}{3N}\left \langle\sum_{\mathbf{k}}\sum_{j=1}^{3N} \mathlarger{\gamma}_{\mathsmaller{V, T}}^{(j)} \right \rangle_{\mathrm{BZ}}$ is the average anharmonicity parameter.

An interesting and compact
result from Eq. \ref{eq: ratio_final} is 
\begin{equation}
    \beta \simeq \beta^{\rm QH} + \frac{6k_{\rm B}}{B_T v}\mathlarger{\overline{\gamma}}_{\mathsmaller{V, T}} \; .
    \label{eq: betaQH}
\end{equation}
This shows that the thermal expansion
differs from that of the QHA 
owing to the mixed second derivative
of the phonon energy of Eq. \ref{AnhParameterDefined}.
In general, 
thermal expansion requires
the consideration of both the 
volume and temperature dependence
of the free energy.

\subsection{Why is the QHA Unreliable for Thermal Expansion?}

Eq.~\ref{eq: betaQH} shows why the quasiharmonic model  predicts a small thermal expansion coefficient when phonon
frequencies have a temperature dependence that varies with volume. 
Only in the limit of no temperature dependence, which means $\mathlarger{\overline{\gamma}}_{\mathsmaller{V, T}}=0$, Eq.~\ref{eq: betaQH} reduces to $\beta^{\rm QH} = \beta$.
Here we estimate some magnitudes
of these effects. 

The change of  internal energy is estimated as
\begin{equation}
    \Delta U_0 = \frac{1}{2}B_0V_0\left(\frac{\Delta V}{V_0}\right)^2\simeq B_0V\cdot (0.01)^2\simeq 10^{0\sim 1} \mathrm{meV} \; , 
\end{equation}
which gives 
\begin{equation}
    B_T V = B_0 V \simeq 10^{4\sim 5} \,\mathrm{meV}=10^{1\sim 2} \,\mathrm{eV}\; .
\end{equation}
Thus,
\begin{equation}
    \frac{6k_{\rm B}}{B_T\beta V} \simeq \frac{6\times 8.617\times 10^{-5} \,\mathrm{eV}\cdot \mathrm{K}^{-1}}{10^{-5\sim -4} \,\rm{K}^{-1}\cdot 10^{1\sim 2} \,\mathrm{eV}}\sim \mathcal{O}\left(10^{-2\sim 0}\right) \,[\mathrm{unitless}] \; ,
\end{equation}
and in some solids, some modes are possible with $0 < \mathlarger{\overline{\gamma}}_{\mathsmaller{V, T}} \sim \mathcal{O}(1)$.

Finally, we have
\begin{equation}
    \beta^{\rm QH}\Big/\beta \sim 1 - \mathcal{O}\left(10^{-2\sim 0}\right)\cdot \mathcal{O}(1)\; .
\end{equation}
Under some circumstances, the  thermal expansion coefficient can be several times larger than the quasiharmonic prediction, which means the QHA fails to get the first-order term for thermal expansion. If the QHA accounts for the leading term of thermal expansion, it is either because 1) the solid is not so anharmonic ($\mathlarger{\overline{\gamma}}_{\mathsmaller{V, T}}$ is small), or 2)
there is a cancellation of positive and negative $\mathlarger{\overline{\gamma}}_{\mathsmaller{V, T}}$ for different phonon
modes despite the anharmonicity.

\subsection{Estimate of Thermal Expansion of NaBr}

We expect these 
$\mathlarger{\gamma}_{\mathsmaller{V, T}}$
to differ for the different phonon modes, but usually all
these parameters $\mathlarger{\gamma}_{\mathsmaller{V, T}}$
are  not available for all phonon modes. 
For NaBr, the energies of acoustic phonons have small changes with temperature compared to the energies of optical phonons. 
We therefore estimate how 
the thermal expansion of NaBr
deviates from the prediction
of the QHA by considering
only  the optical phonons. 

At $T=700\,\rm{K}$, the fractional energy shifts of the LO phonons are estimated by averaging the shifts along the $\Gamma$-$L$ and $\Gamma$-$X$ lines as listed in Table $\textrm{I}$ in the manuscript
\begin{equation}
    \left \langle -\frac{\Delta \omega^{\rm (LO)}}{\omega} \right \rangle_{\mathrm{BZ}} \simeq -\frac{1}{2}[(-0.169)+(-0.132)]=0.15 \; .
\end{equation}
and the anharmonicity parameter by
\begin{equation}
    \mathlarger{\gamma}_{\mathsmaller{V, T}} 
    = -\frac{VT}{\omega}\left ( \frac{\partial^2 \omega}{\partial T\partial V}\right ) 
    \simeq -\frac{VT}{\omega}\frac{\Delta \omega}{\Delta T\Delta V} 
    = -\frac{\Delta \omega}{\omega}\frac{T}{\Delta T} \frac{V}{\Delta V} 
    \simeq -\frac{\Delta \omega}{\omega} \frac{V}{\Delta V} \; .
\end{equation}
With $\displaystyle \frac{a(T=700\,\mathrm{K})}{a(T=10\,\mathrm{K})} = 1.03$, we have $\displaystyle  \frac{V}{\Delta V}=\frac{1.03^3}{1.03^3-1}\simeq 11.78$ and thus $\mathlarger{\overline{\gamma}}^{\mathsmaller{\rm (LO)}}_{\mathsmaller{V, T}}=1.767$. Similarly, we obtain $\displaystyle \left \langle -\frac{\Delta \omega^{(\mathrm{TO}_{1, 2})}}{\omega} \right \rangle_{\mathrm{BZ}} \simeq 0.14$ and $\mathlarger{\overline{\gamma}}^{\mathsmaller{(\mathrm{TO}_{1, 2})}}_{\mathsmaller{V, T}}=1.649$ for the two TO modes. 
The average anharmonicity parameter of NaBr is approximately 
\begin{equation}
    \mathlarger{\overline{\gamma}}_{\mathsmaller{V, T}} = \frac{1}{6}\sum_{i=1}^{6}\mathlarger{\overline{\gamma}}^{\mathsmaller{(i)}}_{\mathsmaller{V, T}} \simeq \frac{1}{6}\left( \mathlarger{\overline{\gamma}}^{\mathsmaller{\rm (LO)}}_{\mathsmaller{V, T}}+2 \mathlarger{\overline{\gamma}}^{\mathsmaller{(\mathrm{TO})}}_{\mathsmaller{V, T}}\right) = 0.844 \; .
\end{equation}

We take the value of $\beta = 3\alpha = 3\times 60.63\times 10^{-6}\,\mathrm{K}^{-1}=182\times 10^{-6} \,\mathrm{K}^{-1}$ \cite{Rao2013}, $B_T=18.5\,\mathrm{GPa}$ \cite{Yosiko1983} and  $a=6.1376\, {\mathrm{\AA}} \,(700\,\mathrm{K})$ \cite{Rao2013}, which gives the volume per atom as $v= a^3/8=2.89\times 10^{-29}\,\mathrm{m}^3$.
So finally, 
\begin{align}
    \beta^{\rm QH}\Big/\beta = 1 - \frac{6k_{\rm B}}{B_T\beta v} \mathlarger{\overline{\gamma}}_{\mathsmaller{V, T}} \simeq 0.28\; ,
\end{align}
which is rather good agreement with the
deficiency of the QHA 
shown in Fig.~\ref{fig: sup_LTC}.

\section{Inelastic neutron scattering experiments}

The INS measurements used a high-purity single crystal of NaBr. 
Crystal quality  was checked by X-ray and neutron diffraction. 
The INS data  were acquired with the time-of-flight Wide Angular-Range Chopper Spectrometer, ARCS, at the Spallation Neutron Source at the Oak Ridge National Laboratory.
The  neutrons had an incident energy of 30\,meV.
The single crystal of [001] orientation  was suspended in an aluminum holder, 
which was  mounted in a closed-cycle helium refrigerator 
for the 10\,K measurement,
and a low-background electrical resistance vacuum furnace for measurements at 300 and 700\,K. 
For each measurement, time-of-flight neutron data 
was collected from 201 rotations of
the crystal in increments of 0.5$^{\circ}$ about the vertical axis.

Data reduction gave the 4D scattering function $S(\mathbf{Q}, \varepsilon)$, where $\mathbf{Q}$ is the 3D wave-vector and $\varepsilon$ is the phonon energy (from the neutron energy loss). 
Measurements with an empty can were  performed to evaluate the background. 
To correct for nonlinearities of the ARCS instrument, 
offsets of the $q$-grid were corrected to first order by fitting 
a set of 45 \emph{in situ} Bragg diffractions, 
which were  transformed to their theoretical positions in the reciprocal space of the NaBr structure. 
The linear  transformation matrix had only a small deviation (less than 0.02) from the identity matrix, showing that the original data had  good quality and the linear correction for $q$-offsets was adequate. 
After subtracting the empty-can background and removing multiphonon scattering calculated 
with the incoherent approximation
(discussed below), 
the intensities from the higher
Brillouin zones  were folded back into an irreducible 
wedge in the first Brillouin zone to obtain the spectral intensities shown in Fig. 2 in the main text.

The multiphonon scattering in the incoherent approximation \cite{Sears1973} is given by
\begin{eqnarray}
    S_{n>1}(\mathbf{Q}, \varepsilon)= \sum_{n=2}^{\infty}\sum_{d}e^{-2W_d}\frac{(2W_d)^n}{n!}\frac{\sigma_{\mathrm{total}, d}}{M_d}A_{n, d}(\varepsilon) \; ,
\label{eq: mps}
\end{eqnarray}
where $\mathbf{Q}$ is the reciprocal space vector, $\varepsilon$ is the phonon energy, and for atom $d\in (\rm Na, Br)$, $\sigma_{\mathrm{total}, d}$ is the total  neutron scattering cross section, $M_d$ is atomic mass, and
\begin{eqnarray}
2W_d=2W_d(| \mathbf{Q}|)= \frac{\hbar ^2|\mathbf{Q}|^2}{2M_d}\int_{0}^{\infty}d\varepsilon \ \frac{g_d(\varepsilon)}{\varepsilon}\coth{\left( \frac{\varepsilon}{2 k_B T}\right )}
\end{eqnarray}
is the Debye-Waller factor.
The $n^{\rm th}$-order partial phonon spectra of atoms $d$ and $\bar{d}$,
$A_{n, d}$ and $A_{n, \bar{d}}$, were calculated as
\begin{align}
    A_{1, d}(\varepsilon) &=\frac{g_d(\varepsilon)}{\varepsilon}\frac{1}{e^{\varepsilon /k_BT} -1} \; ,\\
    A_{1, \bar{d}}(\varepsilon) 
    &=\frac{g_{\bar{d}}(\varepsilon)}{\varepsilon}\frac{1}{e^{\varepsilon /k_BT} -1} \; ,\\
    A_{n, d}(\varepsilon) &= \frac{1}{2}\Big ( A_{1, d}\circledast A_{n-1, d} + \frac{1}{n}A_{1, d}\circledast A_{n-1, \bar{d}}+ \frac{n-1}{n}A_{1, \bar{d}}\circledast A_{n-1, d}\Big )\; , \\
    A_{n, \bar{d}}(\varepsilon) &= \frac{1}{2}\Big ( A_{1, \bar{d}}\circledast A_{n-1, \bar{d}} + \frac{1}{n}A_{1, \bar{d}}\circledast A_{n-1, {d}}+ \frac{n-1}{n}A_{1, {d}}\circledast A_{n-1, \bar{d}}\Big )\; .
\end{align}
Here, $\bar{d}$ refers to the other atom in the unit cell.
The temperature-dependent partial phonon density of states (DOS), $g_d(\varepsilon)$, was obtained by our sTDEP method that used \emph{ab initio} DFT calculations.

\begin{figure}
\centering
\includegraphics[width=\linewidth]{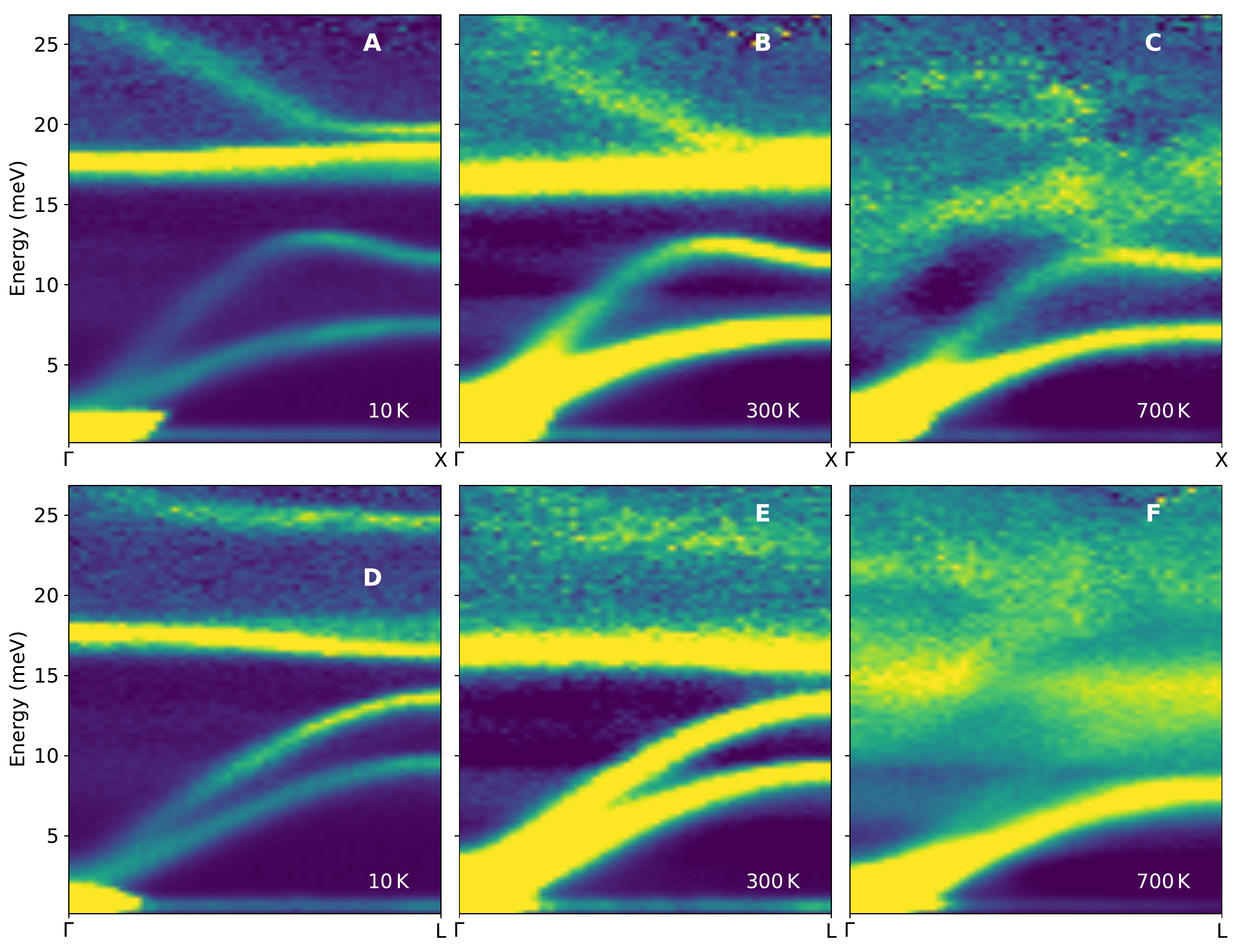}
\caption{{\label{fig: sup_exp}}\textbf{Experimental phonon dispersions of NaBr along $\Gamma$-$X$ and $\Gamma$-$L$.} Phonon dispersions are shown by 2D slices of the $S(\mathbf{Q}, \varepsilon)$ data along the high symmetry lines of $\Gamma$-$X$ \textbf{(a-c)} and $\Gamma$-$L$ \textbf{(d-f)} at the temperature of 10\,K \textbf{(a, d)}, 300\,K \textbf{(b, e)} and 700\,K \textbf{(c, f)}.}
\end{figure}

For NaBr, we truncated Eq.~\ref{eq: mps} at $n=8$, and a global scaling factor was applied to the multiphonon scattering function for normalization. 
Finally, the folded-back data was corrected for the phonon creation thermal factor. This folding technique cancels out the polarization effects and improves the statistical quality by assessing phonon intensities over multiple Brillouin zones.
Fig.~\ref{fig: sup_exp} shows a set of enlarged, separated figures of the scattering data along the $\Gamma \text{-}X$ and $\Gamma \text{-}L$ directions. 

\section{ab initio calculations\label{ab initio calculations}}

\subsection{General}
All DFT calculations were performed with the VASP package using a plane-wave basis set \cite{Kresse1993, Kresse1994, Kresse1996, Kresse1996_2} with  projector augmented wave (PAW) pseudopotentials \cite{Kresse1999} and the Perdew-Burke-Ernzerhof (PBE) exchange correlation functional \cite{Perdew1996}. All calculations used a kinetic-energy cutoff of 550\,meV, a 5$\times$5$\times$5 supercell of 250 atoms, and a 3$\times$3$\times$3 $k$-point grid. Quasiharmonic calculations used PHONOPY \cite{Togo2015}. The sTDEP \cite{Hellman2011, Hellman2013, Hellman2013_2} method was used to calculate anharmonic phonons at elevated temperatures. The Born effective charges and dielectric constants were obtained by DFT calculations in VASP \cite{Gajdos2006}. The non-analytical term of the long-ranged electrostatics was corrected for both quasiharmonic and anharmonic calculations \cite{Gonze1997}. The phonon self-energy was calculated with a 35$\times$35$\times$35 $q$-grid. 

\subsection{Quasiharmonic calculations}
\begin{figure}
    \centering
    \includegraphics[width=\linewidth]{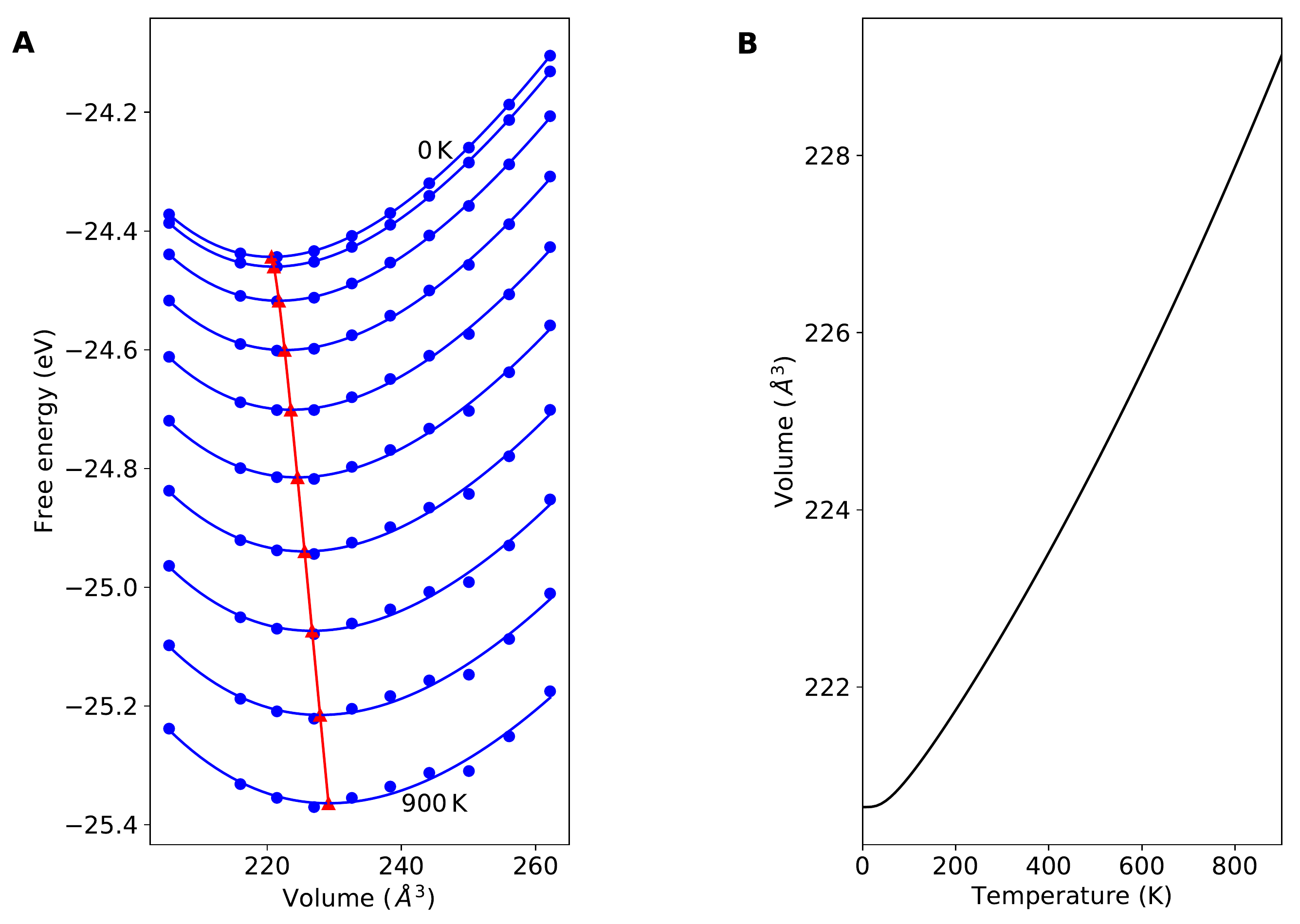}
    \caption{\textbf{Intermediate results for expansion coefficients of NaBr calculated with the QHA.} \textbf{a,} The Helmholtz free energy as a function of temperature and volume. The volume-energy data was fitted to a Birch-Murnaghan equation of state. \textbf{b,} The equilibrium volumes with temperature, obtained by  minimizing  the free energy at each  temperature.}
    \label{fig: sup_qha}
\end{figure}

\begin{figure}
\centering
\includegraphics[width=0.6\linewidth]{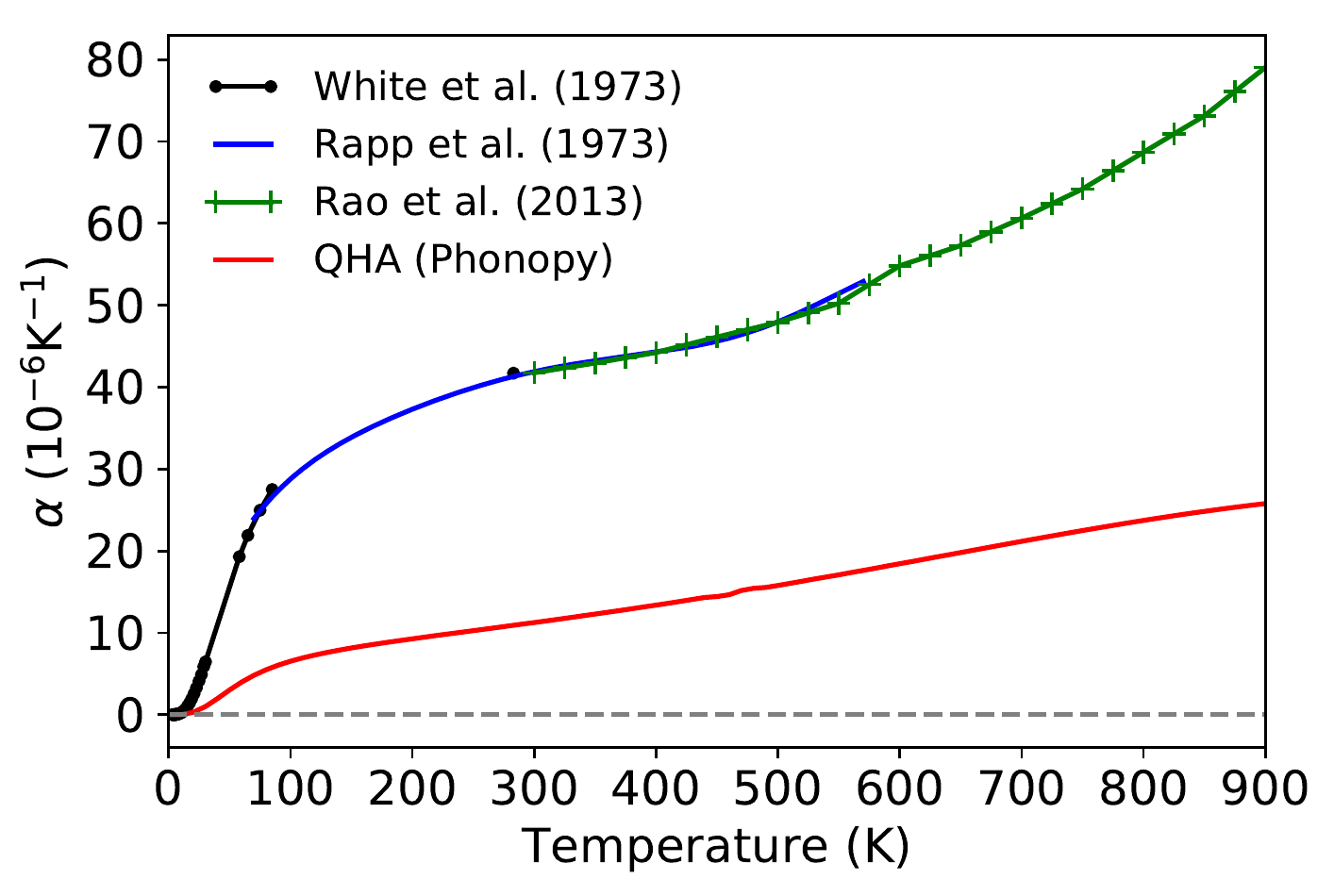}
\caption{{\label{fig: sup_LTC}} \textbf{Thermal expansion coefficients of NaBr, measured and calculated by QHA.} The \emph{ab initio} quasiharmonic predictions (red solid line) are compared to the experimental results \cite{White1973, Rapp1973, Rao2013}. The linear thermal expansion coefficients, $\alpha$, are a factor of four lower than experimental results.}
\end{figure}

The  free energy and the equilibrium volumes calculated with the QHA are shown in Fig.~\ref{fig: sup_qha}. The linear thermal expansion coefficients from measurements and QHA calculations are compared in Fig.~\ref{fig: sup_LTC}. We did not calculate detailed linear thermal expansion coefficients with the stochastically-initialized temperature dependent effective potential method (sTDEP) method, 
but  we compared lattice constants at several temperatures  to illustrate  thermal expansion (see Fig. 1 in the main text). 

\begin{figure}[!htb]
\centering
\includegraphics[width=0.8\linewidth]{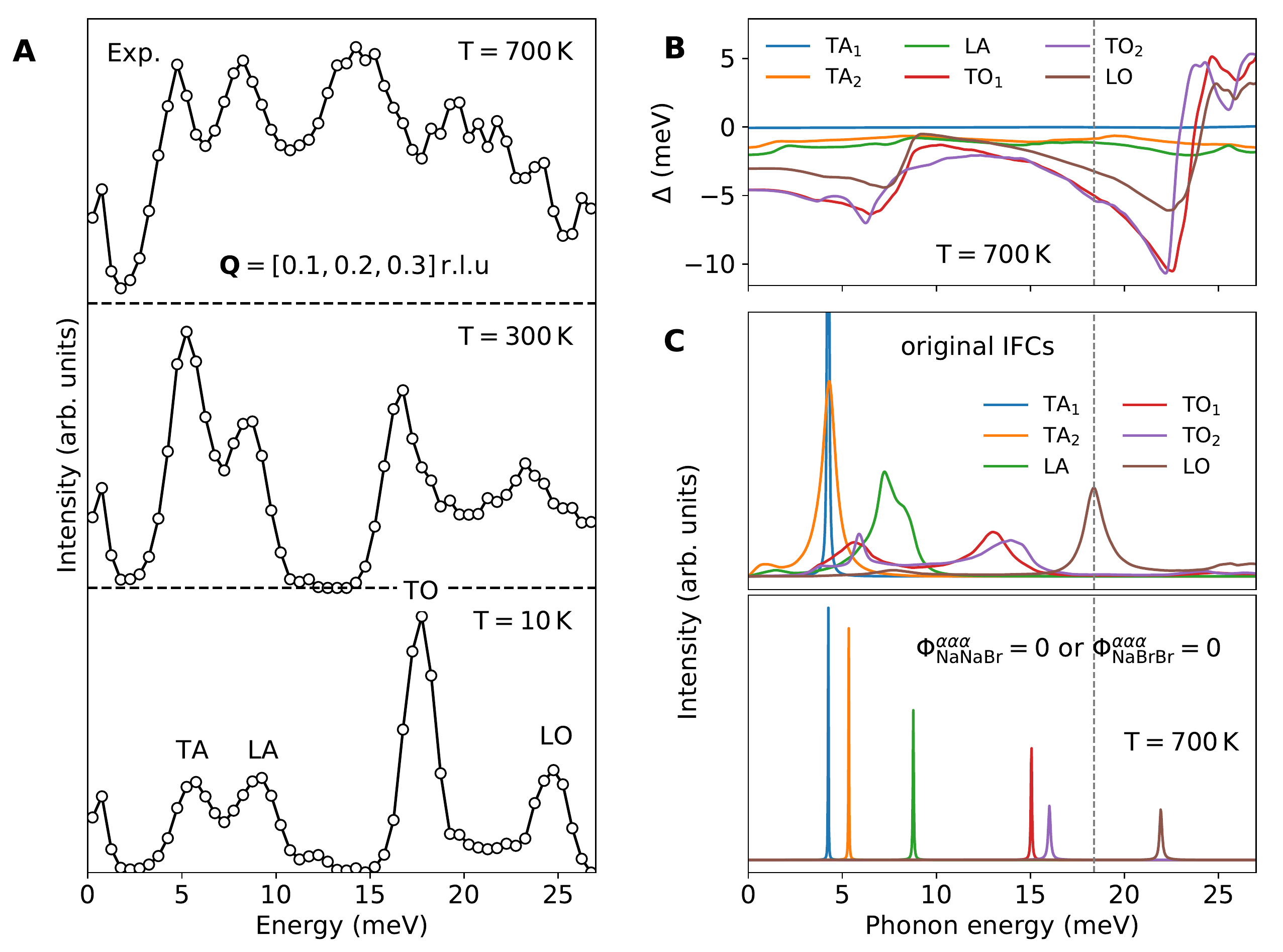}
\caption{{\label{fig: sup_point1}} \textbf{Measured and calculated phonon lineshapes at $\mathbf{Q}=[0.1, 0.2, 0.3]\,\mathrm{r.l.u.}$ and the real part of the phonon self-energy.} \textbf{a,} The 1D cut of $S(\mathbf{Q}, \varepsilon)$ at a constant $\mathbf{Q} = [0.1, 0.2, 0.3]\,\mathrm{r.l.u.}$ (reciprocal lattice units), showing the temperature dependence of phonon lineshapes in NaBr. At this $Q$-point, the LO phonon peak has an energy decrease with temperature of $3\sim 4$\,meV. This  can  be attributed to the real component of the phonon self-energy as shown in \textbf{(b)}.  The intensity data were scaled and offset for clarity. \textbf{c,} By nulling the third-order force constants, $\Phi_{\mathrm{NaNaBr}}^{\alpha \alpha \alpha}$ or $\Phi_{\mathrm{NaBrBr}}^{\alpha \alpha \alpha}$, associated with the nearest-neighbor degenerate triplets, where $\alpha = (x, y, z)$ represents the direction along the Na-Br bond, the lineshapes at this $Q$-point become narrow Lorentzian peaks at 700\,K and the energy decrease of the LO mode vanishes.}
\end{figure}

\begin{figure}[!htb]
\centering
\includegraphics[width=0.8\linewidth]{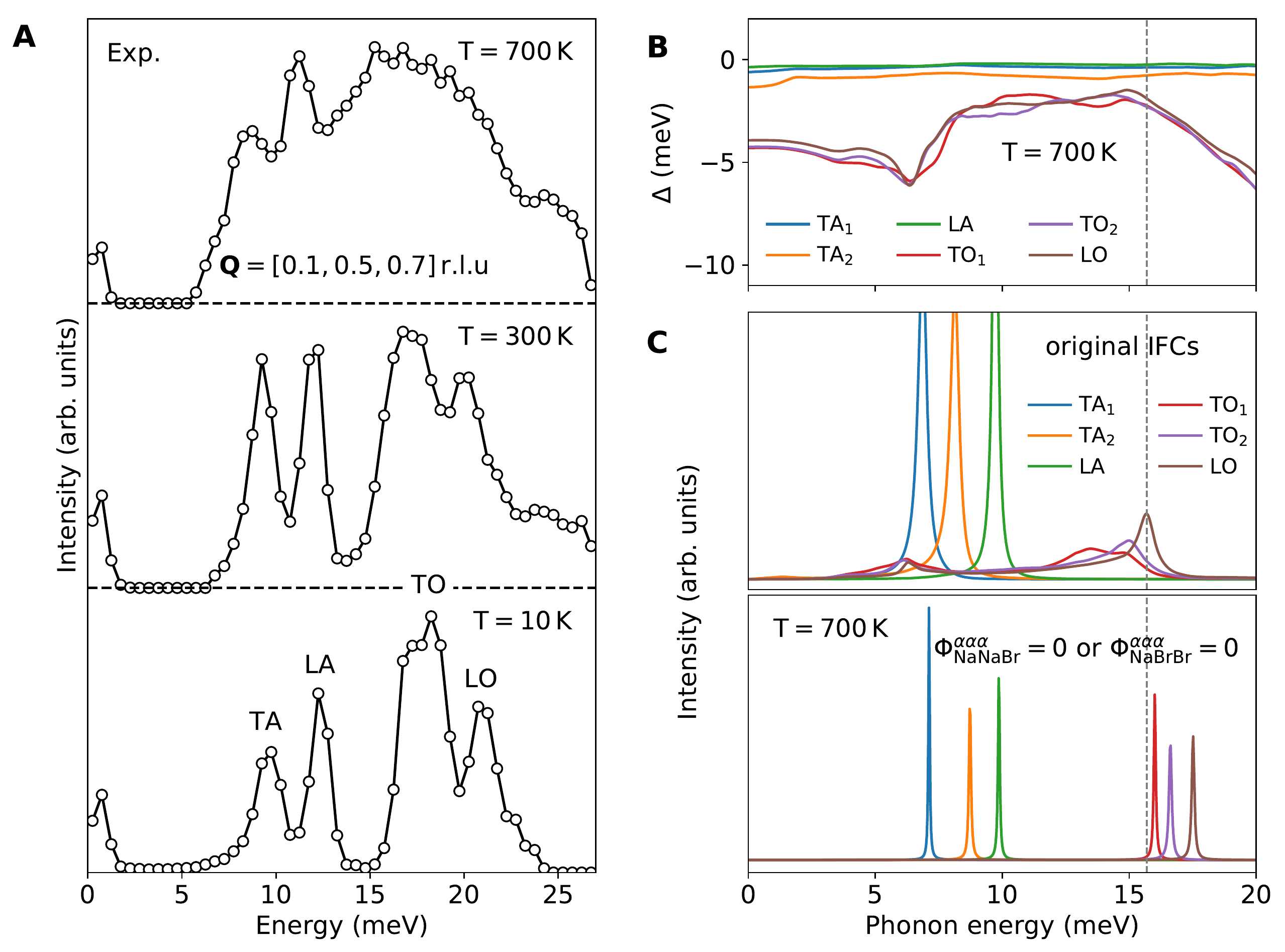}
\caption{{\label{fig: sup_point2}} \textbf{Measured and calculated phonon lineshapes at $\mathbf{Q}=[0.1, 0.5, 0.7]\,\mathrm{r.l.u.}$ and the real part of the phonon self-energy.} The panels are the same quantities in the previous figure, but for $\mathbf{Q}=[0.1, 0.5, 0.7]\,\mathrm{r.l.u.}$ It is seen again that the LO phonon mode shifts to a lower energy at 700\,K, mainly due to the cubic interactions.}
\end{figure}

\subsection{Anharmonic calculations: sTDEP method}

With harmonic forces, the instantaneous position ($\mathbf{u_i}$) and velocity ($\dot{\mathbf{u}}_i$) of the $i$th atom are the sums of contributions from $3N$ normal modes 
\begin{align} 
\mathbf{u}_i = & \sum_{s=1}^{3N} \boldsymbol{\epsilon}_{is} A_{is} \sin( \omega_s t+\delta_s ) \; ,\\ 
\dot{\mathbf{u}}_i = & \sum_{s=1}^{3N} \boldsymbol{\epsilon}_{is} A_{is} \omega_s \cos( \omega_s t+\delta_s) \; ,
\end{align}
where $ A_{s}$ is the normal mode amplitude,  $\delta_s$ is the phase shift, $\omega_s$ and $\boldsymbol{\epsilon}_s$ are eigenvalue and eigenvector corresponding to mode $s$.

To obtain a set of positions and velocities that correspond to a canonical ensemble, we choose the $A_s$ and $\delta_s$ so they are normally distributed around their mean value.  Each mode $s$ should contribute, on average, $k_\mathrm{B} T/2$ to the internal energy. Then
\begin{equation} 
\langle A_{is} \rangle = \sqrt{\frac{\hbar (2n_s+1) }{2 m_i \omega_s}} \approx \frac{1}{\omega_s}\sqrt{\frac{k_\mathrm{B}T}{m_i}}\; , 
\end{equation}
where the approximate result is in the classical limit, $\hbar \omega \ll k_\mathrm{B}T $. The appropriate distribution of atomic positions and velocities are
\begin{align} 
\mathbf{u}_i & = \sum_{s=1}^{3N} \boldsymbol{\epsilon}_{is} \langle A_{is} \rangle \sqrt{-2\ln \xi_1}\sin 2\pi\xi_2 \; , \\ 
\dot{\mathbf{u}}_i & = \sum_{s=1}^{3N} \omega_s \boldsymbol{\epsilon}_{is} \langle A_{is} \rangle \sqrt{-2\ln \xi_1}\cos 2\pi\xi_2 \; ,
\end{align} 
where $\xi_n(n=1, 2)$ represent a uniform distribution of random numbers between $(0, 1)$, which are transformed to a normal distribution using the standard Box-Muller transform \cite{Wallace1972, Shulumba2017}.

In practice, we performed first-principles calculations on a temperature-volume grid covering  five temperatures and five volumes. We chose the five temperatures as $T=\{10,\,300,\,450,\,600,\,700\}$ K and the five volumes linearly spaced within $\pm 5\%$ around the equilibrium volumes. We iterated for 3 to 5 times until the force constants were converged.


Using results from many-body theory, 
the phonon 
frequencies were obtained from the dynamical matrix for the constants $\{ \Phi_{ij} \}$,
and then
corrected by the 
real ($\Delta$) and imaginary ($\Gamma$) parts of the phonon self-energy.
The imaginary part of the 
phonon self-energy was calculated with the third-order force constants,
\begin{align}
\label{eq: CubicSecondOrder}
\Gamma_{\lambda}(\Omega) & =  \frac{\hbar\pi}{16}
\sum_{\lambda'\lambda''}
\left|
\Phi_{\lambda\lambda'\lambda''}
\right|^2
\big{\{}(n_{\lambda'}+n_{\lambda''}+1) \nonumber \\
&\times \delta(\Omega-\omega_{\lambda'}-\omega_{\lambda''})+
(n_{\lambda'}-n_{\lambda''}) \\
&\times \left[
\delta(\Omega-\omega_{\lambda'}+\omega_{\lambda''}) -
\delta(\Omega+\omega_{\lambda'}-\omega_{\lambda''})
\right]
\big{\}} \; , \nonumber
\end{align}
where $\Omega$($=E/\hbar$) is the probing energy.
The real part was obtained by a Kramers-Kronig transformation
\begin{eqnarray}
\Delta(\Omega)=\mathcal{P}\int\frac{1}{\pi}\frac{\Gamma(\omega)}{\omega-\Omega}\mathrm{d}\omega \; .
\label{eq: Kramers-Kronig}
\end{eqnarray}
Equation \ref{eq: CubicSecondOrder}
is a sum over all possible three-phonon interactions, where $\Phi_{\lambda\lambda'\lambda''}$ is the three-phonon matrix element obtained from the cubic force constants $\Phi_{ijk}$ by  Fourier transformation, $n$ is the Bose-Einstein thermal occupation factor giving the number of phonons in each mode, and the delta functions conserve energy and momentum.

\section{Phonons away from  high symmetry lines}

The anharmonicity and its origin with first-neighbor Na-Br bonds are not only true for phonons
along the high-symmetry lines, but for the whole Brillouin zone.
Figures~\ref{fig: sup_point1} and \ref{fig: sup_point2} (similar to Fig.~3 in the manuscript)
show  lineshapes from experiment and computation at two  arbitrary points
in the Brillouin zone, along with the calculated real part of the phonon self-energy. The thermal softening of the LO phonon modes at 700\,K is seen over the Brillouin zone. The real part of the phonon self-energy, arising from cubic anharmonicity to second order, is the main cause of
these thermal shifts and broadenings. 

\providecommand{\noopsort}[1]{}\providecommand{\singleletter}[1]{#1}%